\documentclass[letterpaper,journal]{IEEEtran}
\usepackage{amsmath,amsfonts,amssymb,mathtools}
\usepackage{bm}
\usepackage{algorithmic}
\usepackage{algorithm}
\usepackage{array}
\usepackage[caption=false,font=normalsize,labelfont=sf,textfont=sf]{subfig}
\usepackage{textcomp}
\usepackage{stfloats}
\usepackage{url}
\usepackage{verbatim}
\usepackage{graphicx}
\usepackage{cite}
\usepackage{xcolor}
\usepackage{makecell}
\DeclareMathOperator{\E}{\mathbb{E}}

\DeclareMathOperator{\SURE}{SURE}
\DeclareMathOperator{\diag}{diag}

\DeclareMathOperator*{\argmin}{argmin}

\makeatletter
\let\old@widetilde\widetilde
\def\widetildeto#1#2{\phantom{\widetilde{#2}}\mathllap{\widetilde{\phantom{#1}\mathllap{#2}}}}
\makeatother

\usepackage{hyperref}
\hypersetup{
    colorlinks,
    linkcolor={red!50!black},
    citecolor={blue!50!black},
    urlcolor={blue!80!black}
}

\title{Equivariance2Inverse: A Practical Self-Supervised CT Reconstruction Method Benchmarked on Real, Limited-Angle, and Blurred Data}

\author{Dirk Elias Schut\textsuperscript{1}\thanks{\textsuperscript{1} Computational Imaging Group, Centrum Wiskunde en Informatica (CWI), Science Park 123, 1098 XG Amsterdam, The Netherlands}, Adriaan Graas\textsuperscript{1}, Robert van Liere\textsuperscript{1,2}\thanks{\textsuperscript{2} Visualization Cluster, Eindhoven University of Technology, PO Box 513, 5600 MB Eindhoven, The Netherlands}, Tristan van Leeuwen\textsuperscript{1,3}\thanks{\textsuperscript{3} Mathematisch Instituut, Utrecht University, Budapestlaan 6, 3584 CD Utrecht, The Netherlands}}

\usepackage{tikz}

\newcommand\submittedtexttop{%
\footnotesize © 2026 IEEE. Personal use of this material is permitted. Permission from IEEE must be obtained for all other uses, in any current or future media, including reprinting/republishing this material for advertising or promotional purposes, creating new collective works, for resale or redistribution to servers or lists, or reuse of any copyrighted component of this work in other works.}

\newcommand\submittedtextbottom{%
\footnotesize This article has been accepted for publication in IEEE Transactions on Computational Imaging. This version includes all revisions by the authors based on peer reviewer suggestions, but it has not been fully edited, and content may change prior to final publication. Citation information: DOI \href{https://doi.org/10.1109/TCI.2026.3684416}{10.1109/TCI.2026.3684416}}

\newcommand\submittednoticetop{%
\begin{tikzpicture}[remember picture,overlay]
\node[anchor=north,yshift=-12pt] at (current page.north) {\fbox{\parbox{\dimexpr0.82\textwidth-\fboxsep-\fboxrule\relax}{\submittedtexttop}}};
\end{tikzpicture}%
}
\newcommand\submittednoticebottom{%
\begin{tikzpicture}[remember picture,overlay]
\node[anchor=south,yshift=12pt] at (current page.south) {\fbox{\parbox{\dimexpr1.0\textwidth-\fboxsep-\fboxrule\relax}{\submittedtextbottom}}};
\end{tikzpicture}%
}

\begin{document}
\maketitle
\submittednoticetop
\submittednoticebottom

\begin{abstract}
Deep learning has shown impressive results in reducing noise and artifacts in X-ray computed tomography (CT) reconstruction. Self-supervised CT reconstruction methods are especially appealing for real-world applications because they require no ground truth training examples. However, these methods involve a simplified X-ray physics model during training, which may make inaccurate assumptions, for example, about scintillator blurring, the scanning geometry, or the distribution of the noise. As a result, they can be less robust to real-world imaging circumstances. In this paper, we review the model assumptions of six recent self-supervised CT reconstruction methods. Based on this, we combined concepts of the Robust Equivariant Imaging and Sparse2Inverse methods in a new self-supervised CT reconstruction method called Equivariance2Inverse that is robust to scintillator blurring and limited-angle data. We benchmarked Equivariance2Inverse and the existing methods on the real-world 2DeteCT dataset and on synthetic data with and without scintillator blurring and a limited-angle scanning geometry. The results of our benchmark show that methods that assume that the noise is pixel-wise independent do not perform well on data with scintillator blurring. Moreover, they show that when the distribution of objects is rotationally invariant, this invariance can be used to reduce artifacts in limited-angle reconstructions.
\end{abstract}

\section{Introduction}
\IEEEPARstart{I}{n} X-ray computed tomography (CT), multiple X-ray projection images are combined to form an image representing the inside of an object through a process called image reconstruction. Learned image reconstruction techniques have shown impressive results in reducing noise and artifacts compared to traditional (non-learned) image reconstruction techniques \cite{jin2017deep, adler2018learned}. This is very promising for low-dose (e.g., medical) or high-throughput (e.g., industrial) applications of CT imaging. Learned image reconstruction was first demonstrated using supervised learning. However, supervised learning requires a large dataset of paired input and ground truth data, which can be challenging or expensive to acquire. Unsupervised CT reconstruction methods do not require paired input and ground truth data \cite{ongie2020deep}, making these methods more practical for real-world use.

Several approaches for unsupervised CT reconstruction exist that use different data and training strategies. Diffusion-based methods \cite{song2022solving, chung2022improving, chung2022diffusion, song2023pseudoinverse} learn a prior distribution of the reconstructed volumes, and they have outperformed supervised learning-based methods for several reconstruction problems. However, diffusion-based methods require ground truth data of objects for training, which still makes it challenging to acquire a suitable training dataset. Methods based on implicit neural representations (INRs) \cite{sun2021coil, zang2021intratomo, zha2022naf, wu2023self} train a separate neural network for each scan. INR-based methods are particularly useful when only a single scan of an object is available; however, they are less suitable for high-throughput applications, as training a new network for every scan is computationally intensive. Moreover, these methods do not benefit from the large diversity of image features that large datasets have to offer. INR-based methods are sometimes referred to as self-supervised. However, we will use the term self-supervised exclusively to refer to a different category of methods. Self-supervised methods are trained on a dataset of measurement data from multiple scans, where in every loss function call, data from the same scan is used both as input and as target \cite{hendriksen2020noise2inverse, chen2022robust, unal2024proj2proj, gruber2024sparse2inverse, gruber2025noisier2inverse, sechaud2026equivariant, xu2023rotational}. Measurement data is typically simpler to acquire than ground truth data, making self-supervised methods simpler to train in practice than diffusion-based methods, while offering better performance than INR-based methods. Therefore, this paper focuses on self-supervised methods.

To train a neural network without ground truth data, self-supervised methods rely on an X-ray physics model. Traditional CT reconstruction methods also use an X-ray physics model, which is often highly simplified, for example, assuming a dense-view geometry, a linear projection operator, and additive Gaussian noise \cite{hansen2021computed}. Recent self-supervised CT reconstruction methods have introduced different assumptions, such as a sparse-view geometry \cite{chen2022robust, gruber2024sparse2inverse, sechaud2026equivariant, xu2023rotational}, a non-linear projection operator with Poisson + Gaussian noise \cite{chen2022robust}, and correlated noise \cite{gruber2025noisier2inverse, tachella2024unsure}. The fact that different reconstruction methods make different assumptions raises the question of how well these assumptions reflect real-world data. When the same model assumptions are used for generating data and for evaluating a reconstruction method on that synthetic data, the results may be unrealistically positive. This is known as an inverse crime \cite{wirgin2004inverse, nuyts2013modelling}

In this paper, we investigate how limited-angle geometries \cite{abella2018enabling} and scintillator blurring \cite{graas2025scintillator} affect self-supervised CT reconstruction. These conditions are common in practice, but they are commonly not considered when developing self-supervised CT reconstruction methods. We review the model assumptions of six recent self-supervised CT reconstruction methods. Based on this, we combine concepts of the Robust Equivariant Imaging \cite{chen2022robust} and Sparse2Inverse \cite{gruber2024sparse2inverse} methods in a new self-supervised CT reconstruction method called Equivariance2Inverse that is robust to scintillator blurring and limited-angle data.
Moreover, we benchmark Equivariance2Inverse and the six recent methods to evaluate how their model assumptions affect the reconstruction performance. For this goal, synthetic and real-world data have complementary strengths. Synthetic data can be generated with any X-ray physics model, making it possible to change the model assumptions in isolation \cite{andriiashen2024quantifying, andriiashen2024x}. Real-world data provides a good indication of how a method will perform in practice. Our benchmark uses synthetic data with and without scintillator blurring and a limited-angle geometry to test the robustness of each method to these effects. Moreover, our benchmark uses two datasets of the real-world 2DeteCT dataset \cite{kiss20232detect}.

The structure of the paper is as follows: Section \ref{sec:background} outlines the six recent self-supervised CT reconstruction methods. It first describes common assumptions on X-ray physics, and then describes the different approaches used for self-supervised training. Section \ref{sec:E2I} presents our novel self-supervised CT reconstruction method Equivariance2Inverse (E2I). Section \ref{sec:benchmark} describes our benchmark and section \ref{sec:ablation_study} describes an ablation study. Section \ref{sec:results} describes the results of the benchmark and the ablation study and relates them to the model assumptions of each method. Finally, Sections \ref{sec:discussion} and \ref{sec:conclusion} are the discussion and conclusion. 

\section{Background}
\label{sec:background}
\subsection{Problem formulation}
A CT scanner collects multiple X-ray projection images of an object, and the CT reconstruction algorithm uses this data to create an image of the X-ray attenuation coefficient inside that object. In this paper, 2D objects and 1D detectors are considered. When $n$ X-ray projection images are acquired with a detector that has $m$ pixels, all measurements can be represented by a vector $\bm{y} \in \mathbb{R}^{nm}$. The attenuation coefficient inside the object is discretized into a grid of $j$ by $k$ pixels, represented by a vector $\bm{x} \in \mathbb{R}^{jk}$. A CT reconstruction algorithm is a function $f : \mathbb{R}^{nm} \rightarrow \mathbb{R}^{jk}$, and it performs the task of deriving $\bm{x}$ from the X-ray images $\bm{y}$. However, there may be multiple objects $\bm{x}$ that produce the same measurements $\bm{y}$ because of noise or incomplete measurements. This can be modeled with random variables (which will be notated as bold capital letters): $\bm{X}$ for the objects, and $\bm{Y}$ for the projection data. For a given loss function $l(\cdot)$ the reconstruction method aims to minimize: 
\begin{equation}
    \label{eq:CT_problem}
    \hat{f} = \argmin_f \left(\E\left[l(f(\bm{Y}), \bm{X})\right] \right).
\end{equation}
All expected values in this paper are calculated over all random variables. The joint distribution between $\bm{X}$ and $\bm{Y}$ can be decomposed into the conditional distribution $p(\bm{Y}|\bm{X}=\bm{x})$, and the prior distribution of $\bm{X}$.

\subsection{Forward models}
A forward model approximates the conditional distribution $p(\bm{Y}|\bm{X}=\bm{x})$ by modeling the X-ray physics. All self-supervised methods rely on assumptions related to the forward model, but the assumptions vary between methods, and will be discussed in Section \ref{sec:self-supervised_CT}. A forward model can also be used to generate synthetic CT projection data.

\subsubsection{X-Ray physics}
Here we will explain several aspects of X-ray physics and combine them into a forward model. This model will be used for generating the synthetic benchmark data:
\label{sec:forward_model}
\begin{equation}
    \label{eq:Y_BPG}
    \begin{split}
        &\bm{P} \sim \text{Poisson} \left (\diag({\bm{c}})\exp \left (-A\bm{x} \right ) \right )\\
        &\bm{G} \sim \text{Gaussian}\left(\bm{u}, \diag(\bm{v})\right)\\
        &\bm{Y} = \diag(\bm{w})B\bm{P} + \bm{G}.
    \end{split}
\end{equation}
X-rays are emitted by an X-ray source and they decay exponentially, depending on the local attenuation coefficient of the object $\bm{x}$ they are propagating through. The linear projection operator $A \in \mathbb{R}^{nm \times jk}$ describes how the X-rays traverse through the object, so $\exp \left (-A\bm{x} \right )$ is the absorption for each detector pixel in each projection image. The number of X-ray photons reaching the detector is modeled as a Poisson-distributed random vector $\bm{P}$, due to the quantum nature of X-rays \cite{hansen2021computed}. The mean photon count without attenuation $\bm{c} \in \mathbb{R}^{jk}$ is direction-dependent in cone beam CT scanners, because of the anode heel effect \cite{poludniowski2021spekpy}.

\begin{figure}[!t]
    \centering
    \includegraphics[width=0.9\linewidth]{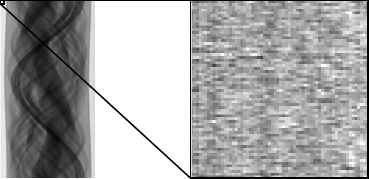}
    \caption{\label{fig:2det_scintillator_blur} Example of scintillator blur in the 2DeteCT dataset. Each horizontal line corresponds to a detector readout at a given time. The zoomed-in region corresponds to constant background radiation, so all variations over time are due to noise. The horizontal correlations in the noise can be attributed to scintillator blurring.}
\end{figure}

Energy-integrating X-ray detectors consist of a scintillator and a sensor layer. The scintillator converts each X-ray photon into multiple visible light photons, and the sensor layer measures the visible light. The conversion from X-ray photons to detector counts can be characterized by a gain $\bm{w} \in \mathbb{R}^{nm}$, which is pixel-dependent \cite{andriiashen2024x}. The visible light photons scatter in the scintillator before reaching the sensor layer, resulting in blurring \cite{gomi2006experimental, howansky2018apparatus, hansen2021computed}. This scintillator blurring can be approximated as a convolution $B \in \mathbb{R}^{nm\times nm}$. Scintillator blurring not only blurs the signal, but also the Poisson component of the noise, resulting in correlated noise (e.g. Figure \ref{fig:2det_scintillator_blur}). Moreover, the electronics in visible light sensors introduce Gaussian noise $\bm{G}$ into their measurements with a pixel-wise variance ($\bm{v} \in \mathbb{R}^{nm}$), and even without any X-rays, there may be a small signal $\bm{u} \in \mathbb{R}^{jk}$ \cite{EMVA1288}. 

\subsubsection{Pre-processing}
In many CT reconstruction methods, pre-processing (also called flatfielding \cite{moy1999does} and log-transforming) is applied to the raw data $\bm{Y}$ to obtain the pre-processed data $\widetilde{\bm{Y}}$:
\begin{equation}
    \label{eq:Y_preprocessing}
    \widetilde{\bm{Y}} = -\log\left(\diag(\bm{p}-\bm{q})^{-1}(\bm{Y}-\bm{q})\right).
\end{equation}

The values of $\bm{p}, \bm{q} \in \mathbb{R}^{nm}$ are obtained using simple calibration measurements. $\bm{p}$ is obtained by averaging multiple measurements with no object in the scanner, and it roughly corresponds to $\bm{c} \odot \bm{w} + \bm{u}$ (where $\odot$ is the element-wise product) in Equation \ref{eq:Y_BPG}. $\bm{q}$ is obtained by averaging multiple measurements with the X-ray source turned off, and it roughly corresponds to $\bm{u} $ in Equation \ref{eq:Y_BPG}.
For $\widetilde{\bm{Y}}$ a linear forward model with additive Gaussian noise with covariance $\Sigma \in \mathbb{R}^{nm\times nm}$ is commonly assumed \cite{buzug2008computed, hansen2021computed}:
\begin{equation}
    \label{eq:Y_simplified}
    \widetilde{\bm{Y}} \sim \text{Gaussian}\left(A\bm{x}, \Sigma\right).
\end{equation}

\subsubsection{The projection operator}
The projection operator $A$ is determined by the scanning geometry, and by the number of acquired projection images. The scanning geometry is how the source, detector, and object move relative to each other, and it is called \textit{complete} when the ranges of motion are sufficient for reconstructing any object within the volume of interest \cite{tuy1983inversion, smith1985image}. Circular geometries that cover an insufficient range of rotations to be complete are called \textit{limited-angle} geometries. Reconstructions from incomplete geometries may have blurring or streaking artifacts. Blurring and streaking artifacts may also appear when the number of projection images is low. Such reconstruction problems are called sparse-view reconstruction problems \cite{gruber2024sparse2inverse, sechaud2026equivariant, crowther1970reconstruction}.

\subsection{Supervised CT reconstruction (SUP)}
Equation \ref{eq:CT_problem} can be interpreted as a supervised deep learning problem. In that setting, the optimization of the reconstruction function $f$ is performed by drawing paired samples from ($\bm{X}, \bm{Y}$), and doing stochastic gradient descent over these samples, which approximates optimizing over the expected value of the loss.

In this paper, we follow the FBPConvNet approach \cite{jin2017deep}, where a neural network $g : \mathbb{R}^{jk} \rightarrow \mathbb{R}^{jk}$ is applied as a post-process to a Filtered Backprojection (FBP) reconstruction \cite{buzug2008computed}. An FBP reconstruction can be represented by a matrix $R \in \mathbb{R}^{jk \times nm}$, and it requires pre-processed projection data $\widetilde{\bm{Y}}$ as input. Together $g$ and $R$ form a learned reconstruction function for pre-processed data: $g(R(\widetilde{\bm{Y}}))$. A mean squared error (MSE) loss is used to optimize the parameters of $g$, resulting in the loss function:
\begin{equation}
\label{eq:supervised}
\E\left [ \left\|g(R\widetilde{\bm{Y}}) - \bm{X}\right\|_2^2 \right ].
\end{equation}
\subsection{Self-supervised CT reconstruction methods}
\label{sec:self-supervised_CT}

\begin{figure*}[!t]
    \centering
    \includegraphics[width=\linewidth]{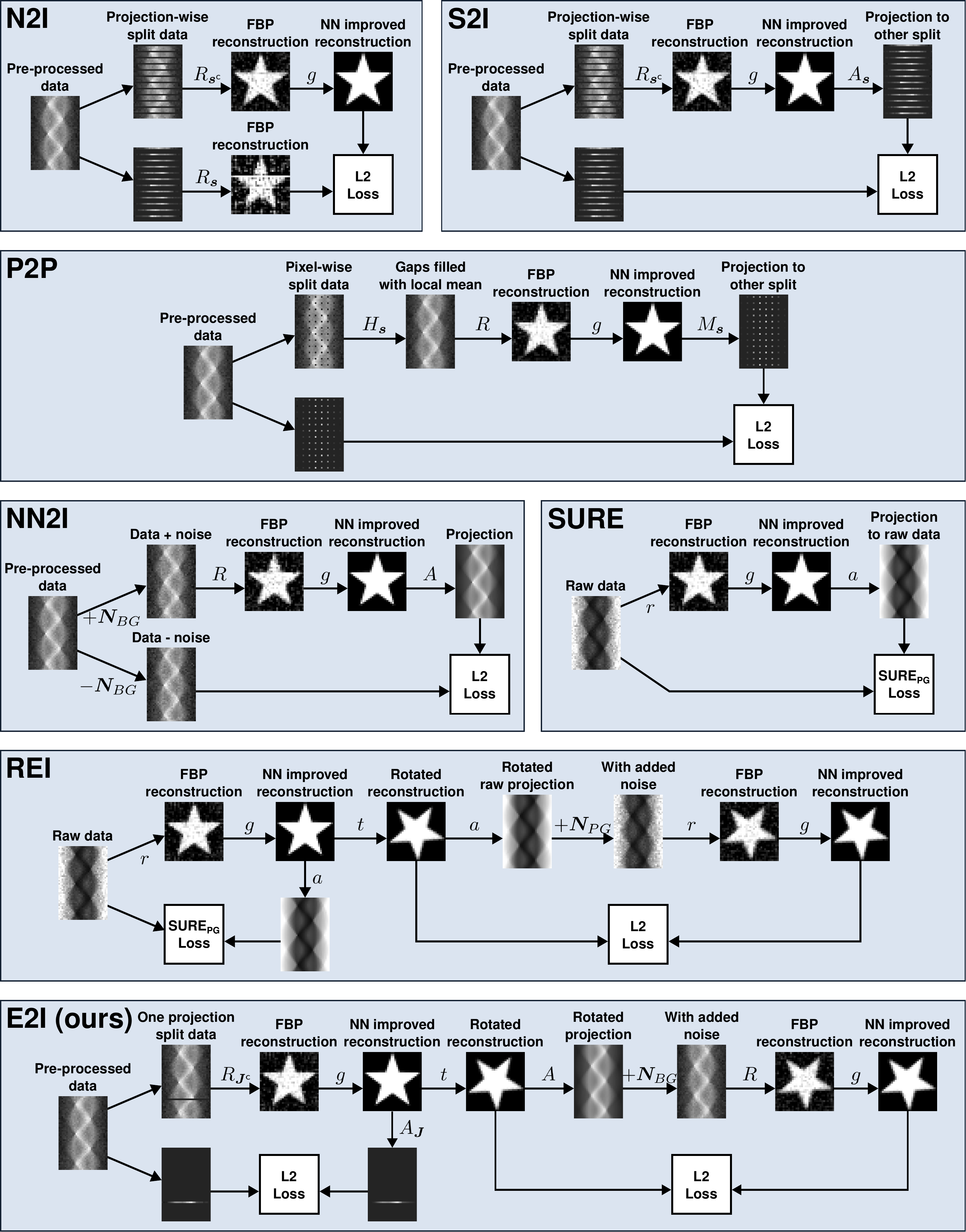}
    \caption{\label{fig:methods} Illustration of how the loss is calculated in the self-supervised CT reconstruction methods. The arrows with letters correspond to function calls or matrix multiplications using the same notation as in Table \ref{tab:existing_methods}.}
\end{figure*}

In this section, we review six recent self-supervised CT reconstruction methods. Unless otherwise mentioned, pre-processed projection data was used as input ($\widetilde{\bm{Y}}$ in Equations \ref{eq:Y_preprocessing} \& \ref{eq:Y_simplified}). The loss functions and the assumptions made by each method are provided in Table \ref{tab:existing_methods}, and an illustration of how the methods are calculated is provided in Figure \ref{fig:methods}.

\begin{table*}[!tb]
\begingroup
\small
\centering
\caption{\label{tab:existing_methods}Loss functions and assumptions of the compared self-supervised CT reconstruction methods.}
\setlength{\tabcolsep}{2mm}
\begin{tabular}{lll}
\textbf{Method} & \textbf{Loss Function} & \textbf{Assumptions}\\\Xhline{1.5\arrayrulewidth}
\begin{tabular}[c]{@{}l@{}}Noise2Inverse (N2I) \end{tabular} &
$\E\left[\frac{1}{4}\sum_{\bm{s}\in{\{\bm{s}_1,..,\bm{s}_4\}}}\left\|g(R_{\bm{s}^{\mathsf{c}}}\widetilde{\bm{Y}}_{\bm{s}^{\mathsf{c}}})-R_{\bm{s}}\widetilde{\bm{Y}}_{\bm{s}}\right \|^2_2 \right ]$ &
\begin{tabular}[c]{@{}l@{}}Pre-processed noise is projection-wise\\ independent and zero mean\vspace{1.2mm}\end{tabular} \\
\begin{tabular}[c]{@{}l@{}}Sparse2Inverse (S2I) \end{tabular} &
$\E\left[\frac{1}{4}\sum_{\bm{s}\in{\{\bm{s}_1,..,\bm{s}_4\}}}\left\|A_{\bm{s}}g(R_{\bm{s}^{\mathsf{c}}}\widetilde{\bm{Y}}_{\bm{s}^{\mathsf{c}}})-\widetilde{\bm{Y}}_{\bm{s}}\right\|^2_2\right]$ & \begin{tabular}[c]{@{}l@{}}Pre-processed noise is projection-wise\\ independent and zero mean\vspace{1.2mm}\end{tabular} \\
\begin{tabular}[c]{@{}l@{}}Proj2Proj (P2P) \end{tabular} &
$\E\left[\frac{1}{16}\sum_{\bm{s}\in{\{\bm{s}_1,..,\bm{s}_{16}\}}}\left\|M_{\bm{s}}Ag(RH_{\bm{s}}\widetilde{\bm{Y}})-M_{\bm{s}}\widetilde{\bm{Y}})\right\|^2_2\right]$ &
\begin{tabular}[c]{@{}l@{}}Pre-processed noise is pixel-wise\\ independent and zero mean\vspace{1.2mm}\end{tabular} \\
\begin{tabular}[c]{@{}l@{}}Noisier2Inverse\\ (NN2I) \end{tabular} &
$\E\left[\left\|Ag(R(\widetilde{\bm{Y}}+\bm{N}_\text{BG}))-(\widetilde{\bm{Y}}-\bm{N}_\text{BG})\right\|^2_2\right]$ &
\begin{tabular}[c]{@{}l@{}}Pre-processed noise is blurred \\ Gaussian with known parameters\vspace{1.2mm}\end{tabular} \\
SURE &
$\E[\SURE_{\text{PG}}(a(g(r(\bm{Y}))), \bm{Y})]$ &
\begin{tabular}[c]{@{}l@{}}Raw noise is Poisson +\\ Gaussian with known parameters\vspace{1.2mm}\end{tabular}\\
\begin{tabular}[c]{@{}l@{}}Robust Equivariant\\Imaging (REI) \end{tabular} &
$\E[\SURE_{\text{PG}}(a(g(r(\bm{Y}))), \bm{Y}) + \lambda h_\text{REI}(\bm{Y})]$ &
\begin{tabular}[c]{@{}l@{}}Raw noise is Poisson + Gaussian with known\\ parameters and $\bm{X}$ is rotation invariant\vspace{1.2mm}\end{tabular}\\
\begin{tabular}[c]{@{}l@{}}Equivariance2Inverse\\ (E2I) \end{tabular}&
$ \E\left[\left\|A_{\bm{J}}(g(R_{\bm{J}^{\mathsf{c}}}(\widetilde{\bm{Y}}_{\bm{J}^{\mathsf{c}}})))-\widetilde{\bm{Y}}_{\bm{J}}\right\|^2_2 + \lambda h_\text{E2I}(\widetilde{\bm{Y}}_{\bm{J}^{\mathsf{c}}}, {\bm{J}^{\mathsf{c}}})\right] $&
\begin{tabular}[c]{@{}l@{}}Pre-processed noise is blurred Gaussian with\\ known parameters and $\bm{X}$ is rotation invariant\end{tabular} \vspace{0.2mm}\\\Xhline{1.5\arrayrulewidth}
\end{tabular}\\
\endgroup
\vspace{1.5mm}
$g$ is the neural network that is optimized. $A$ is the projection operator, and $R$ is an FBP reconstruction. $a(\cdot)$ and $r(\cdot)$ are non-linear versions of $A$ and $R$ including the (inverse) pre-processing of Equation \ref{eq:Y_preprocessing}. $\bm{Y}$ is the raw projection data, and $\widetilde{\bm{Y}}$ is the pre-processed projection data. In N2I, S2I and E2I, $\widetilde{\bm{Y}}_{\bm{s}}$ are only the projection images with indices in $\bm{s}$, and $\widetilde{\bm{Y}}_{\bm{s}^{\mathsf{c}}}$ are all other projection images in $\widetilde{\bm{Y}}$. In P2P the sets $\bm{s}$ represent pixelwise selections, and $H_{\bm{s}}$ is the operator that replaces the pixels in set $\bm{s}$ with their local means and $M_{\bm{s}}$ is a mask that only selects the pixels in $\bm{s}$. $\bm{N}_\text{BG}$ is blurred Gaussian noise that is randomly generated every time the loss function is calculated. $\SURE_{\text{PG}}$ is the SURE loss for a Poisson + Gaussian noise distribution. $h_{\text{REI}}$ and $h_{\text{E2I}}$ are equivariance terms and their definitions can be found in Equations \ref{eq:h_REI} and \ref{eq:h_E2I}, respectively. $\bm{J}$ is the index of one projection image, randomly sampled every time the loss function is calculated.
\end{table*}

\subsubsection{Cross-validation methods}
Cross-validation methods split the projection data into two parts: \textit{the network input data} and \textit{the target data}. The network is trained to predict the target data from the network input data. Different splits are used in different training iterations so that all data is assigned both as network input and as target data. Cross-validation methods require that the noise is independent and zero mean between both parts. The network can learn to approximate the signal of the target data because this information is correlated to the network input data, but it can not learn to predict the noise of the target data because the noise is independent. Therefore, it should converge towards the minimum mean squared error (MMSE) estimator of the noise-free target data from the noisy input data \cite{batson2019noise2self, krull2019noise2void}. After the network has finished training, one approach for inference is to average the reconstructions from multiple input splits \cite{hendriksen2020noise2inverse}. Another approach is to use all data as input during inference. While this approach may introduce some bias, it has shown good experimental performance \cite{hendriksen2021deep, gruber2024sparse2inverse, unal2024proj2proj}, and is less computationally expensive. This second approach was used for all cross-validation methods in the benchmark of this paper.

The main benefits of cross-validation methods are that their required assumptions are often met in practice, and that they do not require knowing the exact noise level. However, a downside is that a more efficient or less biased estimator may be possible because the full data can not be used as the network input data during training.

We compare three cross-validation CT reconstruction methods with slightly different loss functions (see Table \ref{tab:existing_methods}): In \textbf{Noise2Inverse (N2I)} \cite{hendriksen2020noise2inverse} 25\% equally spaced projection images are used as target data, and the remaining projection images are used as input data. The neural network weights are optimized to minimize the MSE between the neural network output and an FBP reconstruction of the target data. \textbf{Sparse2Inverse (S2I)} \cite{gruber2024sparse2inverse} uses the same splits between target and network input data as N2I. However, instead of performing an FBP reconstruction of the target data, the neural network output is projected using matrix $A$, and the MSE is calculated between the projected neural network output and the target data. \textbf{Proj2Proj (P2P)} \cite{unal2024proj2proj} uses pixel-wise instead of projection-wise splitting between network input and target data. The network input data is the projection data with every fourth pixel horizontally and vertically replaced by its local mean. The loss is calculated in the projection domain, like S2I, but only over the pixels that were replaced in the neural network input.

For sparse-view or limited-angle reconstruction problems, N2I may learn to approximate streaking artifacts, because the FBP reconstructions of the target data contain streaking artifacts. S2I was designed to avoid this problem by not performing an FBP reconstruction of the target data. While this approach does not incentivize learning streaking artifacts, the neural network may learn to produce arbitrary components in the null-space of $A$, because adding any null-space component to the neural network output does not affect the loss. Nevertheless, in the experiments of the original S2I paper, S2I consistently outperformed N2I on sparse-view data \cite{gruber2024sparse2inverse}.

Scintillator blurring was not mentioned in the original publications of any of these methods. However, it is expected that blurring will negatively affect the denoising performance of P2P, because blurring introduces correlations in the noise between neighboring pixels, which for P2P violates the requirement that the noise should be independent between the network input and target data. Blurring does not cause correlations between projections, so N2I and S2I should be relatively unaffected. This may explain why N2I was not affected by scintillator blurring when applied to real data \cite{hendriksen2021deep}, while pixel-wise splitting for X-ray denoising was affected \cite{graas2025scintillator}.

\subsubsection{Noisier2Inverse (NN2I)}
In NN2I \cite{gruber2025noisier2inverse}, new noise is generated from a blurred Gaussian distribution, which should approximate the distribution of the noise in the projection data, and the neural network is trained to remove the noise. During training, the neural network is applied to the projection data with new noise added, and an MSE loss is calculated between this value and the projection data with the same noise subtracted (see Table \ref{tab:existing_methods}). This converges to the MMSE estimator of the noise-free projection data from the noisy projection data with additional added noise ($\widetilde{\bm{Y}}+\bm{N}_\text{BG}$). During inference, the network is applied to projection data with no additional added noise, leading to some bias, but producing good results in practice \cite{gruber2025noisier2inverse}.

The main benefit of NN2I is that it is the only method in this section that was designed and tested for cases where correlated noise is present. When correlated noise is assumed, the added noise should simply be correlated in the same way. A downside of this is that it requires estimating the noise correlation and the noise level.

\subsubsection{Stein's Unbiased Risk Estimator (SURE)}
SURE \cite{stein1981estimation} is a function that uses knowledge of the noise model to provide an unbiased estimator of the MSE. Variants of SURE exist for multiple noise models \cite{tachella2024unsure, monroy2025generalized}. However, to the knowledge of the authors, no SURE-based estimator has currently been derived that matches the physics-based forward model of Equation \ref{eq:Y_BPG}.

A variant of SURE exists for Poisson + Gaussian noise with uniform gain $\gamma \in \mathbb{R}$ and standard deviation $\sigma \in \mathbb{R}$ \cite{le2014unbiased}. For a given vector of noise-free projection data $\bm{z} \in \mathbb{R}^{jk}$, the noisy measurement is assumed to be a random variable $\bm{Z} = \gamma\bm{P}+\bm{G}$, with $\bm{P} \sim \text{Poisson}(\bm{z}/\gamma)$ and $\bm{G} \sim \text{Gaussian}\left(\bm{0}, \sigma^2I\right)$. Moreover, let $b : \mathbb{R}^{jk} \rightarrow \mathbb{R}^{jk}$ be a weakly differentiable function. In that case the expectation of the SURE loss equals the expectation of the MSE between the processed noisy measurements $b(\bm{Z})$ and the noisy-free data $\bm{z}$:
\begin{equation}
    \label{eq:SURE_PG}
    \E\left[\SURE_{\text{PG}}(b(\bm{Z}), \bm{Z})\right] = \E\left[\left\|b(\bm{Z})-\bm{z}\right \|^2_2\right].
\end{equation}
This $\text{SURE}_{\text{PG}}$ loss was used for self-supervised CT reconstruction \cite{chen2022robust}, and we refer to that paper on how to calculate $\text{SURE}_{\text{PG}}$. The Poisson + Gaussian assumption pertains to the raw data, whereas the intermediate FBP reconstruction requires pre-processed data. To make this work together, the FBP reconstruction operator $R$ was replaced with a non-linear reconstruction operator $r : \mathbb{R}^{nm} \rightarrow \mathbb{R}^{jk}$, which includes the pre-processing done in Equation \ref{eq:Y_preprocessing}, and the forward operator $A$ was replaced with a non-linear forward operator $a : \mathbb{R}^{jk} \rightarrow \mathbb{R}^{nm}$, which includes the inverse of the pre-processing.

The main benefit of SURE is that it converges towards the MMSE estimator of the noise-free projection data from the noisy projection data, which is very similar to the supervised learning loss. The main downside is that SURE requires modeling of the full forward model and calibration of its model parameters. SURE can be sensitive to calibration errors \cite{tachella2024unsure}.

\subsubsection{Robust Equivariant Imaging (REI)}
The distribution of $\bm{X}$ often contains the same object in multiple orientations. The distribution of $\bm{X}$ is said to be invariant to rotations if for every $\bm{x}$ in the distribution of $\bm{X}$ and every rotation matrix $Q \in \mathbb{R}^{jk\times jk}$, the probabilities of $\bm{x}$ and $Q\bm{x}$ are equal: $\mathbb{P}(\bm{x}) = \mathbb{P}(Q\bm{x})$. REI \cite{chen2022robust} optimizes a loss consisting of the $\text{SURE}_{\text{PG}}$ loss (Equation \ref{eq:SURE_PG}), and an additional equivariance term $\E\left [ \lambda h_\text{REI}\left(\bm{Y}\right) \right ]$:
\begin{equation}
\begin{split}
    \label{eq:h_REI}
    \widetildeto{\bm{Y}}{\bm{X}}_{\!\!\:1} &= t(g(r(\bm{Y})), \bm{T}) \\
    \widetildeto{\bm{Y}}{\bm{X}}_{\!\!\:2} &= g(r(a(\widetildeto{\bm{Y}}{\bm{X}}_{\!\!\:1})+\bm{N}_\text{PG})) \\
    \lambda h_\text{REI}\left(\bm{Y}\right) &= \lambda \left\|\widetildeto{\bm{Y}}{\bm{X}}_{\!\!\:1} - \widetildeto{\bm{Y}}{\bm{X}}_{\!\!\:2}\right\|_2^2
\end{split}
\end{equation}
$g(r(\bm{Y}))$ is the reconstructed image (where $r$ is the non-linear version of $R$). Function $t(\cdot)$ rotates this image by a random amount $\bm{T}$. New projection data is generated from the rotated image by applying the projection operator $a$ and adding noise $\bm{N}_\text{PG}$ with a Poisson + Gaussian distribution, which is assumed to be the distribution of the noise in $\bm{Y}$. From this projection data a new image $\widetildeto{\bm{Y}}{\bm{X}}_{\!\!\:2}$ is reconstructed. Because $\widetildeto{\bm{Y}}{\bm{X}}_{\!\!\:1}$ was rotated, its sparse-view and limited-angle artifacts should be in different positions than in $\widetildeto{\bm{Y}}{\bm{X}}_{\!\!\:2}$. Therefore, optimizing over the MSE between $\widetildeto{\bm{Y}}{\bm{X}}_{\!\!\:1}$ and $\widetildeto{\bm{Y}}{\bm{X}}_{\!\!\:2}$ should reduce these artifacts \cite{chen2021equivariant, tachella2023sensing}. The equivariance weight $\lambda$ specifies the influence of the equivariance term relative to the SURE term.

\section{Equivariance2Inverse (E2I)}
\label{sec:E2I}
Equivariance2Inverse (E2I) is a new self-supervised CT reconstruction method that combines ideas from existing methods, with the goals of being accurate, being simple to calibrate, and being robust to sparsity and correlated noise. Its loss consists of a projection-wise cross-validation term, similar to S2I, and an equivariance term, similar to REI.

\subsection{Cross-validation term}
A projection-wise cross-validation approach similar to S2I is used, because it does not have parameters that require calibration, while still being robust to sparsity and correlated noise. In every iteration of training of E2I, one angle $\bm{J}$ will be randomly sampled, and the projection data from this angle $\widetilde{\bm{Y}}_{\bm{J}}$ will be used as target data. The projection data of all other angles $\widetilde{\bm{Y}}_{\bm{J}^{\mathsf{c}}}$ will be used as input:

\begin{equation}
\label{eq:E2I_CV}
\E \left [ \left \|A_{\bm{J}}(g(R_{\bm{J}^{\mathsf{c}}}(\widetilde{\bm{Y}}_{\bm{J}^{\mathsf{c}}})))-\widetilde{\bm{Y}}_{\bm{J}}\right\|^2_2 \right ].
\end{equation}
The expected value over $\bm{J}$ is the average over all possible values of $\bm{J}$. Therefore, in expectation, this term is the same as the S2I loss with $n$ splits. Like S2I, all data is used as input to the neural network during inference, introducing some bias.

\subsection{Equivariance term}
An equivariance term similar to REI (Equation \ref{eq:h_REI}) is used to reduce limited-angle and sparse-view artifacts. The equivariance term of E2I is based on different forward model assumptions than REI. It assumes that the pre-processed projection data $\widetilde{\bm{Y}}$ has additive blurred Gaussian noise as in Equation \ref{eq:Y_simplified}. While this does not follow the forward model in Equation \ref{eq:Y_BPG} exactly, so it may introduce some bias, the parameters of this forward model are simpler to calibrate (see Sections \ref{sec:calibration} and \ref{sec:calibration_discussion}). The resulting equivariance term is $\E\left [ \lambda h_\text{E2I}\left(\widetilde{\bm{Y}}_{\bm{J}^{\mathsf{c}}}, {\bm{J}^{\mathsf{c}}}\right) \right ]$ with $\lambda h_\text{E2I}\left(\widetilde{\bm{Y}}_{\bm{J}^{\mathsf{c}}}, {\bm{J}^{\mathsf{c}}}\right)$ defined as:
\begin{equation}
\begin{split}
    \label{eq:h_E2I}
    \widetildeto{\bm{Y}}{\bm{X}}_{\!\!\:1} &= t(g(R_{\bm{J}^{\mathsf{c}}}\widetilde{\bm{Y}}_{\bm{J}^{\mathsf{c}}}), \bm{T}) \\
    \widetildeto{\bm{Y}}{\bm{X}}_{\!\!\:2} &= g(R(A\widetildeto{\bm{Y}}{\bm{X}}_{\!\!\:1}+\bm{N}_\text{BG})) \\
    \lambda h_\text{E2I}\left(\widetilde{\bm{Y}}_{\bm{J}^{\mathsf{c}}}, {\bm{J}^{\mathsf{c}}}\right) &= \lambda \left\|\widetildeto{\bm{Y}}{\bm{X}}_{\!\!\:1} - \widetildeto{\bm{Y}}{\bm{X}}_2\right\|_{\!\!\:2}^2
\end{split}
\end{equation}
The equivariance term is calculated from $\widetilde{\bm{Y}}_{\bm{J}^{\mathsf{c}}}$ instead of from $\widetilde{\bm{Y}}$, so that the result of $g(R_{\bm{J}^{\mathsf{c}}}\widetilde{\bm{Y}}_{\bm{J}^{\mathsf{c}}})$ can be re-used from the calculation of the cross-validation term, making the method more computationally efficient.

The equivariance weight $\lambda$ specifies the influence of the equivariance term relative to the cross-validation term. If $\lambda$ is set to a low value, the equivariance term will only have a small effect on the components of $\bm{X}$ in the range-space of $A$. However, it can still improve the estimation of the components of $\bm{X}$ in the null-space of $A$, because these components do not affect the cross-validation term.

\section{Self-supervised CT benchmark}
\label{sec:benchmark}
In this benchmark, the existing self-supervised CT methods were compared with each other, and with E2I, supervised learning (SUP), and an FBP reconstruction.

\subsection{Datasets}
\subsubsection{Synthetic foam datasets}
The goal of using these datasets is to test whether the image quality of the methods is negatively affected by sparsity and blurring and limited-angle geometries. Synthetic data was used so that the exact ground truth and the exact forward model parameters were available. Moreover, the model assumptions could be changed one at a time without affecting the further behavior of the model. A limited-angle and a complete geometry were used, with and without blurring, resulting in four combinations. The noise-free projection data and ground truth volume data of a cylinder of foam were generated using the \texttt{foam\_ct} library \cite{pelt2022foam}. 20 volumes of 256 slices of 256$\times$256 pixels were generated. Two volumes were used for testing, two volumes were used for validation, and the remaining volumes were used for training. 512 projections of width 384 were generated over a range of 180° in a parallel beam geometry. In the complete geometry datasets, all projections were used, and in the limited-angle datasets, the first 256 projections were used, resulting in a 90° missing wedge. The measurement data was generated according to the physics-based forward model in Equation \ref{eq:Y_BPG}, with a constant photon count $\bm{c}$ of 500, a Gaussian variance $\bm{v}$ of $50$, and $\bm{u} = 0$ and $\bm{w} = 1$. In the blurred datasets, $B$ is a convolution with a Gaussian kernel with a standard deviation of 0.8, as was used in \cite{andriiashen2024x}, and in the blurring-free datasets, $B$ is the identity matrix.

\subsubsection{Real-world 2DeteCT datasets}
The goal of using these datasets is to provide a good indication of how well the methods perform in real-world applications. The 2DeteCT dataset \cite{kiss20232detect} was used, which consists of images of a cardboard tube filled with dried fruits, nuts, and lava stones. The overall shape and contrast of the 2DeteCT data approximate those of a medical abdominal scan \cite{kiss20232detect}, and the individual fruits and nuts have natural variation in shape and texture similar to human organs. Raw 2D fan-beam projection data is available, with every image acquired in three modes: (1) high-noise, (2) low-noise, and (3) no filtering (for testing beam hardening). Data from two of these modes was used as two benchmark datasets. The mode 1 (high-noise) data was used with a complete operator. To limit GPU memory use, the projection images in this dataset were downscaled by a factor of two, and every second projection image was used. The mode 2 (low-noise) data was used with a sparse-view and limited-angle projection operator. This operator used 136 equally spaced projections over a range of 136°, which is similar to a low-cost C-arm acquisition \cite{abella2018enabling}.

For both datasets, an FBP reconstruction of all mode 2 data (3600 projections) was used as reference for supervised learning and for calculating error metrics. While this is not a perfect ground truth, it does address the main sources of errors in the 2DeteCT benchmark datasets: The downscaled high-noise benchmark dataset has significantly higher noise, because it uses a 30 times lower tube power than the reference \cite{kiss20232detect}. The limited-angle sparse-view benchmark dataset has roughly 26.5 times fewer projection images than the reference dataset.

During the acquisition of 2DeteCT, the detector of the CT scanner was replaced. Only data from the second detector was used to ensure that the forward model parameters are consistent for all scans. The data from four randomly sampled scanning sessions (200 slices) were used for testing, and four other random sessions were sampled as validation data. The remaining 1770 slices were used for training.

\subsection{Implementation}
\subsubsection{Neural network training} A separate neural network was trained for each combination of method and dataset. The same U-Net architecture \cite{ronneberger2015u} was used in all methods, except that the depth and number of channels were selected based on the image resolution of each dataset to limit the GPU memory use. They were chosen so that at the maximum depth, the resolution and number of channels of the layers were roughly the same between the datasets. On the limited-angle 2DeteCT dataset, the network depth was 7, and the number of channels in the first layer was 8. On the complete 2$\times$ downscaled 2DeteCT dataset, the network depth was 6, and the number of channels in the first layer was 16. On the foam datasets, the network depth was 4, and the number of channels in the first layer was 64.

The optimizer was the ADAM optimizer \cite{kingma2017adammethodstochasticoptimization} with a learning rate of 0.01 and no weight decay. The batch size was 4, which was achieved by parallel training on 4 GPUs (4x Nvidia TITAN X 12GB, 4x Nvidia GTX 1080Ti 11GB, or 4x Nvidia RTX 2080Ti 11GB). Training was stopped after 1000 epochs or when no improvement was observed on the validation loss for 250 epochs. The network weights with the best validation loss were used for inference. PyTorch \cite{paszke2017automatic} and PyTorch Lightning \cite{falcon_2024_10779019} were used for the training, and the projection operator $A$ was implemented in a differentiable and matrix-free manner using Tomosipo \cite{hendriksen2021tomosipo}.

\subsubsection{Forward model parameter calibration} \label{sec:calibration} To estimate the blur convolution kernels used in NN2I and E2I, the approach from \cite{graas2025scintillator} was used. On the foam data, 1024 images with the same image content but with independent noise were generated for this task. On the 2DeteCT data, a background region with no attenuation of 300 sinograms was used. To estimate the standard deviation of the noise, the projection data was first deconvolved, and then the pixel-wise standard deviation was calculated over the same data.

SURE and REI assumed Poisson + Gaussian noise on raw data. On the synthetic data, the exact gain and Gaussian standard deviation were used. On the 2DeteCT data, the gain was estimated by pixel-wise dividing the variance and mean over 300 sinograms of a background region with no attenuation, and then averaging the pixel-wise results. The Gaussian standard deviation was assumed to be zero.

For E2I and REI, networks were trained with multiple power-of-ten values of the equivariance weight $\lambda$ on each combination of method and dataset. The network results with the lowest PSNR on the validation set are reported as the benchmark results.

\subsection{Metrics}
The mean and standard deviation of the PSNR and the Structural Similarity Index Measure (SSIM) \cite{wang2004image} over the images of the test set were calculated as evaluation metrics. The PSNR is inversely related to the supervised learning loss (Equation \ref{eq:supervised}). The SSIM predicts the perceived quality by a human observer.

\section{E2I noise model ablation study}
\label{sec:ablation_study}
In this ablation study, the contribution of the noise modeling in the equivariance term (Equation \ref{eq:h_E2I}) was investigated. Two modifications of E2I were tested: In the \textit{E2I (no eq. blur)} modification, non-blurred Gaussian noise was added instead of blurred Gaussian noise in the equivariance term. In the \textit{E2I (no eq. noise)} modification, no noise was added in the equivariance term.

These variants of E2I were compared on the blurred and non-blurred limited-angle foam datasets, and both 2DeteCT datasets. The standard deviation of the non-blurred Gaussian noise was calculated on the same data as the standard deviation of the blurred Gaussian noise in Section \ref{sec:calibration}, but the deconvolution of the projection data was not performed. The same procedures as in the benchmark were used to train and evaluate the neural networks, including optimizing $\lambda$ over factors of ten based on the validation-set PSNR.

\section{Results}
\label{sec:results}
\begin{figure}[t]
    \centering
    \includegraphics[width=\linewidth]{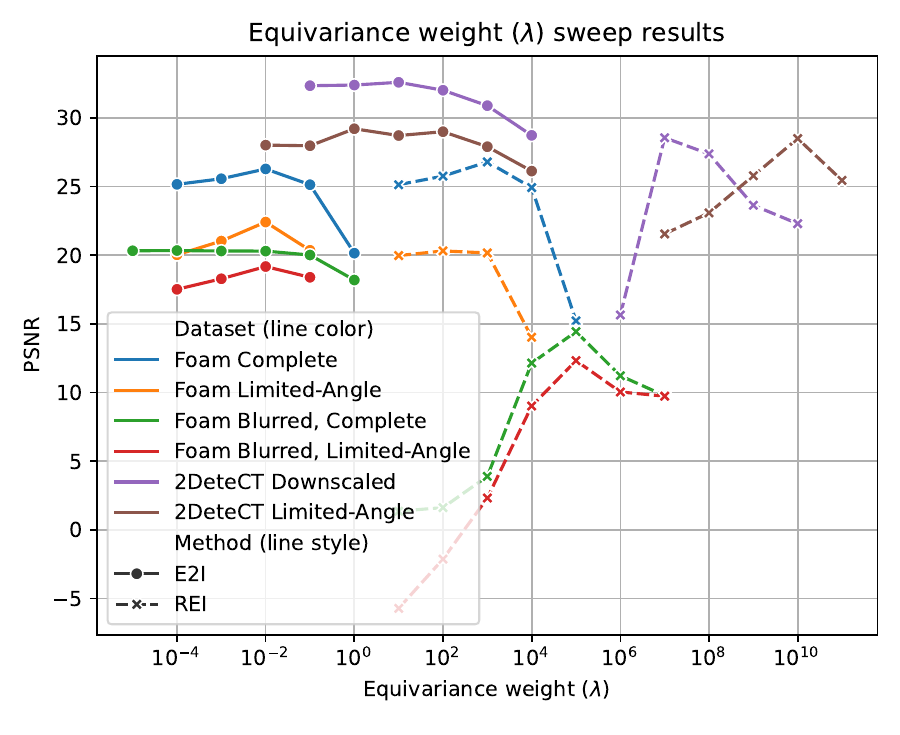}
    \caption{\label{fig:parameter_sweep} The test-set PSNR of neural networks trained with different values of the equivariance weight $\lambda$. A different line color was used for each dataset, and a different line style for each method.}
\end{figure}

\begin{figure*}[t]
    \centering
    \includegraphics[width=\linewidth]{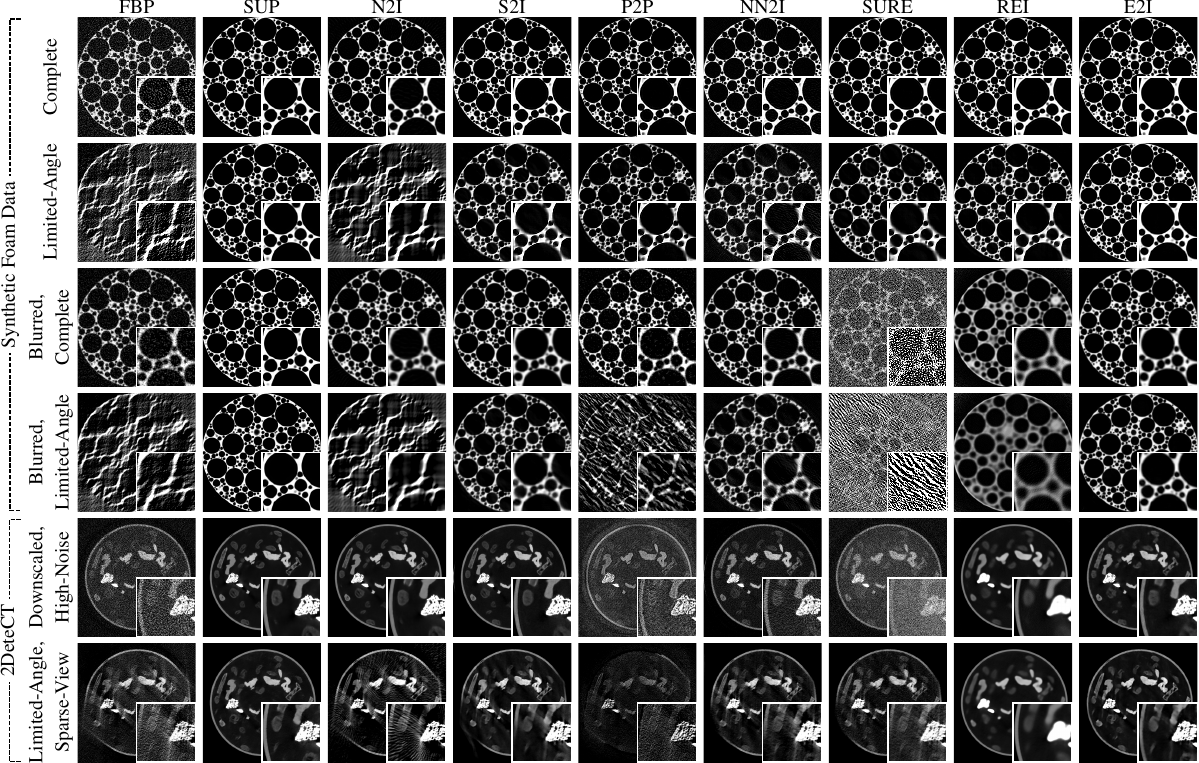}
    \caption{\label{fig:results_grid} A reconstruction from each method on the first image of the test set of each dataset. The insets provide a two times magnified view of the center-left side of the objects. The inset on the 2DeteCT data shows a lava stone (white), some dried fruits or nuts (grey), and the edge of the cardboard tube (near the left edge), which in the limited-angle reconstructions may be incomplete because of limited-angle artifacts.}
\end{figure*}

The benchmark results on the synthetic data and the 2DeteCT data are shown in Tables \ref{tab:metrics_foam} and \ref{tab:metrics_2detect}, respectively. Figure \ref{fig:parameter_sweep} shows the PSNR for all tested values of the equivariance weight $\lambda$. The ablation study results are shown in Table \ref{tab:ablation_study}. An example of a reconstruction of each method on each dataset is shown in Figure \ref{fig:results_grid}.

\begin{table*}[thb]
\begingroup
\small
\centering
\caption{\label{tab:metrics_foam}Benchmark results on the synthetic data.}
\setlength{\tabcolsep}{1.5mm}
\begin{tabular}{l|ll|ll|ll|ll|}
           & \multicolumn{2}{c|}{Complete}                           & \multicolumn{2}{c|}{Limited-Angle} & \multicolumn{2}{c|}{Blurred, Complete} & \multicolumn{2}{c|}{Blurred, Limited-Angle} \\
           & PSNR                          & SSIM                  & PSNR              & SSIM             & PSNR        & SSIM        & PSNR             & SSIM            \\\hline
FBP&13.72 ± 0.09 & 0.43 ± 0.01 & 7.31 ± 0.14 & 0.19 ± 0.01 & 16.69 ± 0.11 & 0.46 ± 0.01 & 9.65 ± 0.17 & 0.22 ± 0.01\\
SUP&29.89 ± 0.36 & 0.99 ± 0.00 & 23.36 ± 0.46 & 0.96 ± 0.00 & 28.95 ± 0.36 & 0.99 ± 0.00 & 22.19 ± 0.48 & 0.92 ± 0.00\\\hline
N2I&24.90 ± 0.16 & 0.86 ± 0.00 & 9.75 ± 0.18 & 0.23 ± 0.01 & 18.72 ± 0.12 & 0.81 ± 0.00 & 10.45 ± 0.19 & 0.23 ± 0.01\\
S2I&25.48 ± 0.19 & 0.95 ± 0.00 & 18.95 ± 0.20 & 0.72 ± 0.01 & \underline{\textbf{20.46 ± 0.13}} & \underline{\textbf{0.91 ± 0.00}} & \textbf{17.32 ± 0.24} & \textbf{0.67 ± 0.01}\\
P2P\textsuperscript{1}&21.72 ± 0.20 & 0.91 ± 0.01 & 17.87 ± 0.45 & 0.77 ± 0.01 & 16.78 ± 0.61 & 0.41 ± 0.01 & 8.32 ± 0.34 & 0.13 ± 0.01\\
NN2I&24.10 ± 0.45 & 0.74 ± 0.02 & 17.06 ± 0.36 & 0.50 ± 0.01 & 20.27 ± 0.17 & 0.79 ± 0.02 & 16.44 ± 0.26 & 0.51 ± 0.01\\
SURE\textsuperscript{1}&25.74 ± 0.21 & \textbf{0.96 ± 0.00} & 19.36 ± 0.21 & 0.73 ± 0.01 & 1.40 ± 0.04 & 0.12 ± 0.00 & -7.82 ± 0.17 & 0.03 ± 0.00\\
REI\textsuperscript{1,2}&\underline{\textbf{26.80 ± 0.19}} & \underline{\textbf{0.96 ± 0.00}} & \textbf{20.32 ± 0.26} & \textbf{0.81 ± 0.01} & 14.44 ± 0.27 & 0.60 ± 0.01 & 12.33 ± 0.27 & 0.37 ± 0.01\\
E2I\textsuperscript{2}&\textbf{26.29 ± 0.22} & 0.93 ± 0.00 & \underline{\textbf{22.42 ± 0.38}} & \underline{\textbf{0.92 ± 0.00}} & \textbf{20.35 ± 0.13} & \textbf{0.89 ± 0.00} & \underline{\textbf{19.18 ± 0.23}} & \underline{\textbf{0.86 ± 0.01}}\\\hline
\end{tabular}\\
\endgroup
\vspace{1.5mm}
The best results are shown in underlined boldface, and the second bests in boldface. The methods marked with \textsuperscript{1} assume that the noise is pixel-wise independent. The methods marked with \textsuperscript{2} use an equivariance loss term, and the table shows the results for the value of $\lambda$ with the highest PSNR.
\end{table*}
\begin{table}[t]
\begingroup
\small
\centering
\caption{\label{tab:metrics_2detect}Benchmark results on 2DeteCT.}
\setlength{\tabcolsep}{1.3mm}
\begin{tabular}{l|ll|ll|}
           & \multicolumn{2}{c|}{\begin{tabular}[c]{@{}c@{}}2$\times$ Downscaled,\\ Complete, High-Noise\end{tabular}}& \multicolumn{2}{c|}{\begin{tabular}[c]{@{}c@{}}Limited-Angle, Sparse-\\View, Low-Noise\end{tabular}} \\
           & PSNR                          & SSIM                  & PSNR              & SSIM             \\\hline
FBP & 16.39 ± 0.53 & 0.05 ± 0.00 & 17.00 ± 0.49 & 0.07 ± 0.00\\
SUP & 33.67 ± 0.69 & 0.78 ± 0.01 & 30.37 ± 0.63 & 0.59 ± 0.02\\\hline
N2I & \underline{\textbf{33.66 ± 0.69}} & \underline{\textbf{0.78 ± 0.01}} & 23.77 ± 1.13 & 0.30 ± 0.04\\
S2I & 28.60 ± 1.21 & 0.64 ± 0.03 & 28.05 ± 0.75 & 0.46 ± 0.02\\
P2P\textsuperscript{1} & 17.55 ± 0.34 & 0.08 ± 0.00 & 17.10 ± 0.35 & 0.06 ± 0.01\\
NN2I & 17.71 ± 1.41 & 0.08 ± 0.02 & 28.03 ± 0.81 & \textbf{0.49 ± 0.02}\\
SURE\textsuperscript{1} & 5.45 ± 1.43 & 0.00 ± 0.00 & 21.09 ± 0.39 & 0.09 ± 0.01\\
REI\textsuperscript{1,2} & 28.56 ± 0.62 & 0.58 ± 0.01 & \textbf{28.50 ± 0.81} & 0.48 ± 0.02\\
E2I\textsuperscript{2} & \textbf{32.60 ± 0.67} & \textbf{0.69 ± 0.02} & \underline{\textbf{29.21 ± 0.71}} & \underline{\textbf{0.51 ± 0.02}}\\\hline
\end{tabular}\\
\endgroup
\vspace{1.5mm}
The best results are shown in underlined boldface, and the second bests in boldface. The methods marked with \textsuperscript{1} assume that the noise is pixel-wise independent. The methods marked with \textsuperscript{2} use an equivariance loss term, and the table shows the results for the value of $\lambda$ with the highest PSNR.
\end{table}

\begin{table*}[t]
\centering
\begingroup
\small
\centering
\caption{\label{tab:ablation_study}E2I noise model ablation study results.}
\setlength{\tabcolsep}{1.3mm}
\begin{tabular}{l|ll|ll|ll|ll|}
               & \multicolumn{2}{c|}{Foam Limited-Angle} & \multicolumn{2}{c|}{\begin{tabular}[c]{@{}c@{}}Foam Blurred,\\ Limited-Angle\end{tabular}} & \multicolumn{2}{c|}{\begin{tabular}[c]{@{}c@{}}2DeteCT 2x Downscaled,\\ Complete, High-Noise\end{tabular}} & \multicolumn{2}{c|}{\begin{tabular}[c]{@{}c@{}}2DeteCT Limited-Angle,\\ Sparse-View, Low-Noise\end{tabular}} \\
               & PSNR & SSIM & PSNR & SSIM & PSNR & SSIM & PSNR & SSIM \\ \hline
E2I      & 22.42 ± 0.38 & 0.92 ± 0.00 & 19.18 ± 0.23 & 0.86 ± 0.01 & 32.60 ± 0.67 & 0.69 ± 0.02 & 29.21 ± 0.71 & 0.51 ± 0.02\\
E2I (no eq. blur) & N.A. & N.A. & 19.04 ± 0.23 & 0.86 ± 0.01 & 32.58 ± 0.68 & 0.67 ± 0.02 & 29.36 ± 0.69 & 0.51 ± 0.02\\
E2I (no eq. noise) & 21.84 ± 0.35 & 0.92 ± 0.00 & 18.09 ± 0.25 & 0.76 ± 0.01 & 32.58 ± 0.68 & 0.68 ± 0.02 & 29.34 ± 0.70 & 0.51 ± 0.03\\ \hline
\end{tabular}
\endgroup \\
\vspace{1.5mm}
The E2I results are the same as the corresponding results in Tables \ref{tab:metrics_foam} and \ref{tab:metrics_2detect}.
\end{table*}

\subsection{Blurring and noise model assumptions}
S2I, P2P, NN2I, and SURE calculate the loss in the projection domain, but make different noise model assumptions. We will compare these four methods to show the effects of these noise model assumptions. The synthetic data without blurring (left two columns of Table \ref{tab:metrics_foam}) matches the noise model assumptions of SURE exactly; in this case, SURE should be an unbiased estimator, which explains why it performs best among these four methods. When scintillator blurring is simulated (right two columns of Table \ref{tab:metrics_foam}), the noise is no longer pixel-wise independent, so the methods that assumed pixel-wise independence perform worse. Of the four methods, S2I achieves the second-highest PSNR on the synthetic datasets without blurring and the highest PSNR on the synthetic datasets with blurring. On the 2DeteCT data, blurring is present (Figure \ref{fig:2det_scintillator_blur}), which could explain why SURE had a lower PSNR than S2I, and NN2I. However, it does not explain the low performance of SURE on the Complete, high-noise, downscaled data (Table \ref{tab:metrics_2detect}), because the downscaling reduces the effects of blurring. Other causes may be calibration errors or unmodeled effects, such as scattering. S2I again showed the highest PSNR of these four methods (Table \ref{tab:metrics_2detect}).

\subsection{Limited-angle data and the equivariance loss term}
S2I is similar to N2I, but it should be less affected by $A$ having a non-trivial null space \cite{gruber2024sparse2inverse}. This explains why the difference between the performance (PSNR and SSIM) of N2I and S2I is much bigger on the synthetic limited-angle data than on the synthetic complete data (Table \ref{tab:metrics_foam}, both with and without blurring). The equivariance term of REI was designed to make SURE more robust to $A$ having a non-trivial null space \cite{chen2022robust}. The effect of the equivariance term can be seen on the limited-angle 2DeteCT reconstructions (bottom row of Figure \ref{fig:results_grid}), where only the self-supervised methods with an equivariance term (REI and E2I) correctly reconstructed the tube as a continuous circle. In all our experiments, REI outperformed SURE, so the equivariance term may also be beneficial for complete geometries. On the blurred synthetic data, it appears that the equivariance term compensates to some degree for the incorrect noise-model assumptions of the SURE loss term, because on that data the PSNR and SSIM of REI are much higher than those of SURE (Table \ref{tab:metrics_foam}), and the resulting images are a lot less noisy (Figure \ref{fig:results_grid}). Nevertheless, multiple other methods had better results on the same data. The E2I loss consists of a loss term similar to S2I and an equivariance term. On the Blurred, Complete synthetic data, the performance of S2I was slightly higher than that of E2I, but on all other datasets in our benchmark, E2I performed better.

\subsection{The effect of the equivariance weight}
Figure \ref{fig:parameter_sweep} shows the test-set PSNR of the networks for the REI and E2I methods that were trained with different values of $\lambda$. The value of $\lambda$ with the highest PSNR was the same on the validation set as on the test set in all cases, so the highest point of every curve in Figure \ref{fig:parameter_sweep} corresponds to the benchmark results in Tables \ref{tab:metrics_foam} and \ref{tab:metrics_2detect}. The fact that the optimal value of $\lambda$ was generally higher for REI than for E2I can be explained by the fact that the data consistency term of REI is calculated using the raw data, while the data consistency loss of E2I is calculated on pre-processed data, which typically has lower values. When comparing the optimal value of $\lambda$ with the adjacent power of ten values, the decrease in PSNR is generally larger for REI than for E2I, showing that E2I is less sensitive to tuning $\lambda$.

\subsection{E2I noise model ablation study}
The results in Table \ref{tab:ablation_study} show that on the foam datasets, adding noise in the equivariance term improves the PSNR and SSIM, but the difference between adding blurred or unblurred noise is very small. On the 2DeteCT datasets, the PSNR and SSIM of the three E2I variants are very similar. These results suggest that the distribution of the added noise in the E2I equivariance term affects the results only slightly.

\subsection{Computational costs}
Table \ref{tab:computational_costs} shows the number of calls to the neural network $g$. Additionally, it shows the computation time per iteration and the GPU memory use during training on the synthetic Foam, Blurred, Complete dataset. The N2I, S2I, P2P, and NN2I methods all use the same number of neural network calls and require a similar amount of GPU memory. The use of an equivariance term in the loss adds one additional neural network call, which increases the computation time and GPU memory use. SURE is calculated using a Monte Carlo-based estimate of the divergence term \cite{chen2022robust}, which requires three additional calls to the neural network, strongly increasing the computation time and GPU memory use. When summing up the time used on different computers, the total training time of all neural networks in this paper is approximately eleven months.

\begin{table}[tbh]
\begingroup
\small
\centering
\caption{\label{tab:computational_costs} The computational costs of training the benchmarked methods.}
\begin{tabular}{llll}
\textbf{Method} & \textbf{NN Calls} & \begin{tabular}[c]{@{}l@{}}\textbf{Time per} \\ \textbf{Iteration} (ms) \end{tabular} & \begin{tabular}[l]{@{}l@{}}\textbf{GPU Memory}\\\textbf{Use} (MiB) \end{tabular} \\\hline
N2I    & 1        & 87.4              & 912           \\
S2I    & 1        & 88.0              & 912           \\
P2P    & 1        & 121.6              & 916           \\
NN2I   & 1        & 72.5              & 918           \\
SURE   & 4        & 236.8              & 1828           \\
REI    & 5        & 292.2             & 2144           \\
E2I    & 2        & 132.4              & 1233           \\\hline
\end{tabular}\\
\endgroup
\vspace{1.5mm}
The GPU memory use and computation time per iteration are measured for a batch size of one per GPU on four Nvidia Titan X GPUs on the blurred and complete synthetic foam dataset.
\end{table}

\section{Discussion}
\label{sec:discussion}
\subsection{The importance of calibration}
\label{sec:calibration_discussion}
An inherent difficulty of applying a method to a new dataset is finding the best parameter values for running the method. Moreover, none of the methods specified a calibration approach for their model parameters. On the synthetic foam data, the data generation parameters were used for SURE and REI, and extensive additional calibration measurements were generated to estimate the parameters for NN2I and E2I. Therefore, we expect that the results on the generated data are not strongly affected by calibration errors. On the 2DeteCT data, no exact parameters or calibration measurements were available. Therefore, calibration inaccuracies may have had a larger impact on these results.

REI and SURE depend on the parameters of a Poisson + Gaussian noise model that would have required many additional calibration measurements to estimate accurately \cite{andriiashen2024x, jezierska2011approach}. The Poisson component of X-ray detector noise is typically much larger than the Gaussian component \cite{ding2016modeling}, which led us to configure these methods with $\sigma=0$. The calibration for the parameters of E2I and NN2I on 2DeteCT was done using a background region with no attenuation, so no additional calibration measurements were required. The recently presented UNSURE method \cite{tachella2024unsure} proposes to optimize the model parameters of SURE-type optimizers alongside optimizing the neural network, removing the need for calibration.

\subsection{Extending the forward model}
There is currently no consensus among self-supervised CT reconstruction methods on what forward model assumptions to make. This raises the question whether more aspects of X-ray physics should be modeled. 

Beam hardening \cite{buzug2008computed} is a common artifact, so it would be interesting future work to study how self-supervised CT reconstruction methods are affected by it, and if it could be corrected by self-supervised learning. The 2DeteCT dataset contains the mode 3 data that was acquired specifically for benchmarking beam hardening reduction \cite{kiss20232detect}.

Scintillator blurring could be modeled in more detail by taking into account that it is slightly angle dependent \cite{freed2010fast}. Moreover, the NN2I and E2I methods take into account that scintillator blurring results in correlated noise, but they do not account for the fact that the signal component ($\exp(-A\bm{x})$ in Equation \ref{eq:Y_BPG}) is also blurred, which led to slightly blurry reconstructions on the synthetic blurred data (Figure \ref{fig:results_grid}).

The focal spot of the X-ray source \cite{mohan2020saber} and scattering \cite{andriiashen2024quantifying} may also cause blurring. However, both of these effects only cause blurring of the signal component and not of the noise, so they can not explain the correlated background noise in Figure \ref{fig:2det_scintillator_blur}. Scattering is also material dependent \cite{andriiashen2024quantifying}, and the radius of the blur is much larger than that of scintillator blurring \cite{andriiashen2024quantifying, seibert2005x, malusek2008calculation}.

The modeling of this paper assumes that an energy-integrating detector is used, which is currently the most widely used type of X-ray detector in CT scanners. Photon-counting detectors, which are increasingly being used in CT scanners \cite{greffier2025photon, wehrse2021photon}, do not have a scintillator and therefore do not exhibit scintillator blurring. Photon-counting detectors might exhibit other effects that could require separate modeling, but that is considered outside of the scope of this paper.

\subsection{Estimation using equivariance}
The main aim of the equivariance terms in REI and E2I is to make the network learn how to estimate the components of $\bm{X}$ that are in the null-space of $A$, which should reduce blurring and streaking artifacts in limited-angle or sparse-view reconstruction. Necessary and sufficient conditions on the operator $A$ and the invariance of the distribution of $\bm{X}$ for learning to estimate the null-space components using self-supervised learning were formulated in \cite{tachella2023sensing}. No guarantees about the bias or statistical consistency of REI-like equivariance terms have been derived, but they have been used successfully in several inverse problems \cite{chen2021equivariant, chen2022robust, sechaud2024equivariance, wang2024perspective}.

The recently introduced Equivariant Splitting (ES) self-supervised reconstruction method \cite{sechaud2026equivariant} has been proven to converge to the supervised learning loss of linear inverse problems with an invariant data distribution even when the projection operator $A$ has a non-trivial null-space. It works by using the invariance to interpret data from one operator as being collected from multiple operators \cite{millard2023theoretical}. It would be interesting future work to combine aspects of ES and E2I into a method that is both robust to scintillator blurring and an unbiased estimator of the null-space components.

\section{Conclusion}
\label{sec:conclusion}
The benchmark in this paper evaluated recent self-supervised CT reconstruction methods on synthetic data with and without scintillator blurring and a limited-angle geometry, and on two real-world datasets from 2DeteCT. REI, which is SURE with an additional equivariance term, had a better performance (PSNR and SSIM) than SURE on all benchmark datasets (Tables \ref{tab:metrics_foam} \& \ref{tab:metrics_2detect}). SURE makes strong model assumptions (pixel-wise independent Poisson + Gaussian noise with known parameters), and it was the best-performing method without an equivariance term on the non-blurred synthetic data, where these assumptions were met exactly. However, on the other datasets, where these assumptions were not met exactly, SURE was outperformed by multiple other methods. S2I, on the other hand, had the most general model assumptions (projection-wise independent zero-mean noise), and it performed best or second best of the methods without an equivariance term on all benchmark datasets. The E2I method introduced in this paper combines the robustness of S2I with the performance increase of the equivariance term of REI. The PSNR of E2I was the best or a close second-best (at most 1.06 lower) on all benchmark datasets.

\section{Code and data availability}
The code is available on Github at: \href{https://github.com/D1rk123/equivariance2inverse}{https://github.com/D1rk123/equivariance2inverse}.
The synthetic foam data is available on Zenodo \cite{schut_2025_16735632} at: \href{https://zenodo.org/records/16735632}{https://zenodo.org/records/16735632}. The 2DeteCT dataset \cite{kiss20232detect} is available on Zenodo at: \href{https://zenodo.org/records/8014758}{https://zenodo.org/records/8014758}.

\section*{Acknowledgments}
We gratefully acknowledge our colleagues: Allard Hendriksen for introducing us to the topic of self-supervised CT reconstruction, and Marcos Obando for the useful discussions and feedback.

\bibliographystyle{IEEEtran}
\bibliography{bibliography.bib}

\begin{IEEEbiography}[{\includegraphics[width=1in,height=1.25in,clip,keepaspectratio]{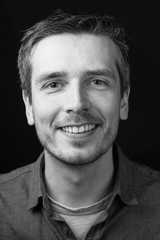}}]{Dirk Elias Schut} is a PhD researcher at the Computational Imaging group of the Centrum Wiskunde \& Informatica (CWI) in the Netherlands. His research is part of the UTOPIA project on bringing per product CT imaging for quality control to food processing factories. He received a Bsc. in Computer Science and a double Msc. in Computer Science and Electrical Engineering, both from Delft University of Technology.\end{IEEEbiography}

\begin{IEEEbiography}[{\includegraphics[width=1in,height=1.25in,clip,keepaspectratio]{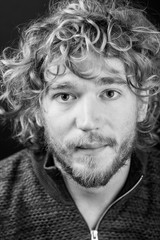}}]{Adriaan Graas} received an MSc. Mathematics from Utrecht University, and is a Ph.D. researcher in the Computational Imaging group at CWI, in Amsterdam. His research, on the topic of Mathematics and Algorithms for 3D Imaging of Dynamic Processes, is conducted within the NDNS+ cluster of mathematics research in the Netherlands. \end{IEEEbiography}

\begin{IEEEbiography}[{\includegraphics[width=1in,height=1.25in,clip,keepaspectratio]{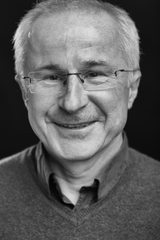}}]{Robert van Liere}
received the PhD degree in computer science from the University of Amsterdam. He is a principal investigator at the Computational Imaging group of the Centrum Wiskunde \& Informatica (CWI) and emeritus full professor at the Eindhoven University of Technology. His research interests include interactive visualization, virtual environments, and human-computer interaction.\end{IEEEbiography}

\begin{IEEEbiography}[{\includegraphics[width=1in,height=1.25in,clip,keepaspectratio]{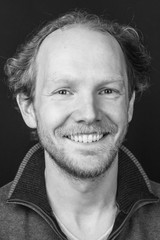}}]{Tristan van Leeuwen} is the group leader of the Computational Imaging group at the Centrum Wiskunde \& Informatica (CWI) and a full professor at Utrecht University. He received his BSc. and MSc. in Computational Science from Utrecht University. He obtained his PhD. in geophysics at Delft University in 2010 and was a postdoctoral researcher at the University of British Columbia in Vancouver, Canada and the CWI. His research interests include: inverse problems, computational imaging, tomography and numerical optimization.\end{IEEEbiography}

\end{document}